\documentclass[prl,twocolumn]{revtex4}
\usepackage{enumerate}
\usepackage{amsfonts,amssymb,amsmath}
\usepackage[]{graphics,graphicx,epsfig}
\usepackage{amsthm}
\usepackage{marvosym}

\usepackage{svg}
\usepackage{fdsymbol}
\usepackage{graphicx}
\usepackage{dcolumn}
\usepackage{natbib}
\usepackage{color}
\usepackage{xcolor}
\usepackage{multirow}
\usepackage{ulem}
\newcommand{\ii}{\\\indent}
\newcommand{\ket}[1]{|{#1}\rangle}
\newcommand{\bra}[1]{\langle{#1}|}

\begin{document}


\title{Quantum Walks in Weak Stochastic Gauge Fields}

\author{Jan W{\'o}jcik}
\affiliation{Institute of Spintronics and Quantum Information, Faculty of Physics, Adam Mickiewicz University, 61-614 Pozna\'n, Poland}
\date{\today}


\begin{abstract}
The behaviour of random quantum walks is known to be diffusive. Here we study discrete time quantum walks in weak stochastic gauge fields. In the case of position and spin dependent gauge field, we observe a transition from ballistic to diffusive motion, with the probability distribution becoming Gaussian. However, in contradiction to common belief, weak stochastic electric gauge fields reveal the persistence of Bloch oscillations despite decoherence which we demonstrate on simulations and prove analytically. The proposed models provide insights into the interplay between randomness and coherent dynamics of quantum walks.
\end{abstract}

\maketitle


Quantum walks introduced by Aharonov, Davidowich and Zagury \cite{Aharonov1993} are simple platforms allowing studies of complex quantum mechanical phenomena. They describe the dynamics of quantum particles on a lattice and thus simulate various physical systems \cite{Cedzich2013,Cedzich2019,Chandrashekar2008,Crespi2013,Jolly2023,Wojcik2004,Zhang2016}. Due to their simplicity they are powerful tools while studying for example topological phenomena \cite{Harper2020,Kitagawa2012,Kitagawa2010,Asboth2012,Asboth2014,Lam2016,Obuse2011,Obuse2015,Grudka2023}. \ii
This paper will focus on the 1D discrete time quantum walks (see \cite{Wu2019} for review). DTQWs have been studied extensively in various experimental setups such as ion traps \cite{Schmitz2009,Zaehringer2010}, NMR \cite{Du2003,Ryan2005}, optical lattices \cite{Dadras2018,Karski2009} and linear optics \cite{Cardano2015,Do2005,Perets2008,Tang2018}. \ii
The general idea behind devising quantum walk was to quantize the random walk model. Thus the DTQW consists of two unitary operations namely the Step operator and the Coin-toss operator. Contrary to diffusive dynamics of classical random walk, the DTQW spread is ballistic. This effect opened up a new brunch of quantum algorithms \cite{Portugal2013,Childs2009,Lovett2010,Singh2021,Buluta2009,Shenvi2003}. \ii
\begin{figure}[t]
    \includegraphics[width=0.48\textwidth]{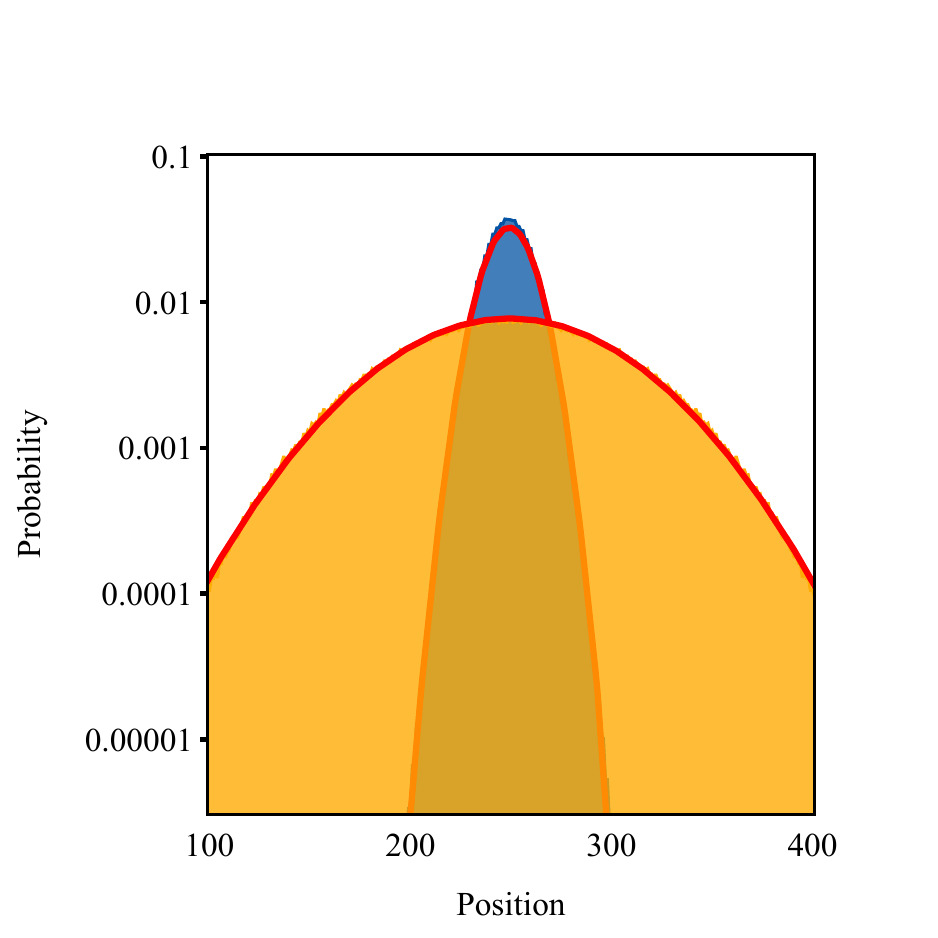}
    \caption{Probability distribution after 400 steps of DTQW on a $d = 501$ vertex cycle in strong ($\phi = 200\frac{2\pi}{d}$ stochastic gauge field averaged over 1000 realizations. The orange distribution corresponds to DTQW in position and spin dependent gauge field, and the blue distribution corresponds to DTQW in only position dependent gauge field. The line of parabolic fit is shown in the figure which suggests the Gaussian probability distribution of the state.}
    \label{binomy}
\end{figure}
\begin{figure}[t]
    \includegraphics[width=0.48\textwidth]{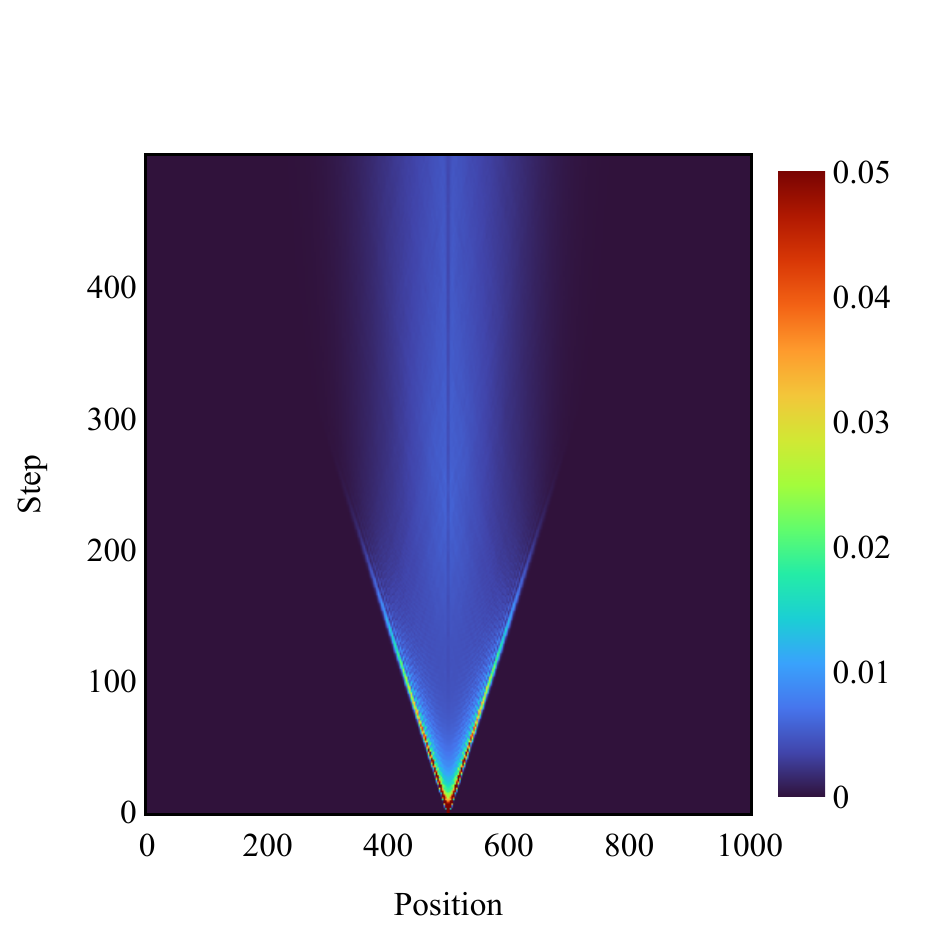}
    \caption{Averaged over 1000 realizations evolution of DTQW in weak (position and spin dependent) gauge field defined in Eq. \ref{ev} of initial state given by Eq. \ref{init} with $r = 0.9$, and $\phi = \frac{2\pi}{d}$.}
    \label{dec}
\end{figure}
Despite the enormous attention the algorithmic usage of DTQW got, Buerschaper and Burnett in 2004 proposed an extension to the model. They introduced the third unitary operation the Phase operator. This simple extension brought back treating DTQW as a platform that can be used to explain complex physical phenomena. This new model of phase quantum walks (PQW) was shortly proved to be useful. Wójcik et al. \cite{Wojcik2004} using the PQW managed to explain unintuitive Bouwmeester et al. \cite{Bouwmeester1999} experimental observations.\ii
Proper setting of the phase operator can lead to correspondence with different physical setups. Thus the model of PQW sparked the interest of many \cite{Arnault2016a,Arnault2016,Boada2017,Bru2016,Buarque2021,Cedzich2016,Cedzich2019,Cedzich2020,Cedzich2013,DiMolfetta2014a,Jay2021,MarquezMartin2018,Sajid2019,Sajid2021,Upreti2020,Yalcinkaya2015}. For example in the \cite{Wojcik2004} Wójcik et al. were investigating the PQW with phase linearly dependent on the position. This model corresponds to an electron in a constant electric field and thus what they found was the phenomenon known as the Bloch osculations. Moreover it was shown that the 1D PQW simulates the dynamics of Dirac particles in arbitrary electric \cite{Debbasch2012,Genske2013,Cedzich2013} and gravitational fields \cite{DiMolfetta2013,DiMolfetta2014,Succi2015}.\ii
In this work, we shall consider DTQWs with random gauge fields where the randomness is time-like. In models with spatial-like randomness one expects to observe Anderson localization. However, when the randomness is time-like the expected effect is decoherence and diffusive behavior of particles. In DTQWs, it was observed e.g. in models with random (in time) coins \cite{Ahlbrecht2011} (for more examples see \cite{KENDON2007,Vieira2014,Brun2003,Joye2011}). The randomness can also be introduced into DTQW by using gauge fields with varying in time magnitudes. Molfetta and Debbasch introduced this idea in a 2018 paper \cite{DiMolfetta2014a} where they discussed DTQWs in random electric and gravitational gauge fields. They found that due to the randomness introduced to the system, the walk loses coherence and the behavior of the walks becomes diffusive. \ii
In this paper we analyze and compere two models of DTQWs in weak stochastic gauge fields. Two mentioned models although similar will differ in their gauge fields. Both gauge fields will be generated by phase factor linearly dependent on position. The difference is that the first model will have phase factor dependent not only on the position but also on the spin whereas the second model will have phase factor only dependent on the position. The second model is thus describing a particle in electric field and we shall refer to it as an electric quantum walk. In the strong field regime of the both models the randomness of the field leads to qualitatively common behavior i.e. the diffusion (see Fig. \ref{binomy}). However, although in the strong field of those models, dynamics seem similar, we show a striking difference between them in the weak field regime.\ii
We investigate the evolution of quantum particle on a $d$ vertex cycle. The state of the particle we will describe as a $2d$ dimensional vector in the Hilbert space. Where $d$ comes from the $d$ possible positions and a factor $2$ comes from the coin degree of freedom. The coin is responsible for choosing whether the particle goes forward or backward. The most general DTQW evolution is governed by a unitary evolution operator 
\begin{equation}
    U = (I^{(d)}\otimes C)\ S,
\end{equation}
where
\begin{equation}
    S \ket{x,\pm1} = \ket{x\pm 1,\pm1}
\end{equation}
is the Step operator and
\begin{equation}
    C = e^{-i\theta \sigma_y},
\end{equation}
\begin{figure}[b]
    \includegraphics[width=0.48\textwidth]{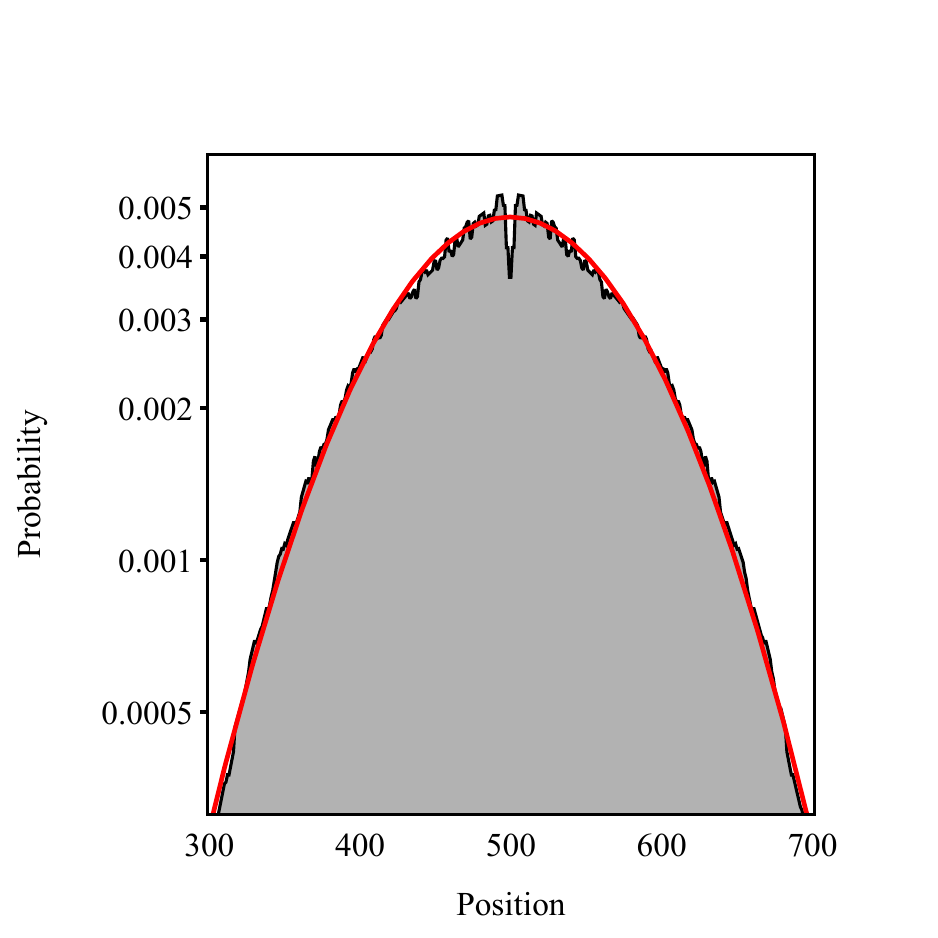}
    \caption{Probability distribution after 400 steps of evolution defined in \ref{ev} averaged over 1000 realizations. The whole evolution is shown in Fig. \ref{dec}. The line of parabolic fit is shown in the figure which suggests the Gaussian probability distribution of the state.}
    \label{state}
\end{figure}
is the Coin-toss operator which in general can be an arbitrary $\mathcal{U}(2)$ matrix but we choose this specific one for simplicity ($\sigma_y$ being the Pauli matrix). In this form, the Coin-toss operator is just a rotation of two state coin around the $y$ axis. For the simplicity, we put $\theta = \pi/4$. As we mentioned we will use two different gauge fields. The first will be generated by a position and spin dependent phase shift operator
\begin{equation}\label{phib}
    \Phi_B \ket{x,\pm} = e^{\pm i\phi x}\ket{x,\pm},
\end{equation}
where the $\phi$ is the magnitude of the field. Since we discuss the walk on a cycle $\phi = \frac{2\pi}{d}q$ but we focus on the weak field regime so we will stick to $\phi = \frac{2\pi}{d}$ ($q=1$). The PQW with this particular gauge field has been already discussed \cite{Grudka2023a}. However, we propose a setup in which the field is applied with probability $r$. Thus the state is evolved in the following way
\begin{equation}\label{ev}
    \ket{\Psi(t+1)}=\left\{
                \begin{array}{ll}
                   U\ket{\Psi(t)}~~~~~~~~~~\text{with probability $1-r$},\\
                   U\Phi_{B}\ket{\Psi(t)}~~~~~~\text{with probability $r$}.
                \end{array}
              \right.
\end{equation}
Let's evolve the following initial state
\begin{equation}\label{init}
    \ket{\Psi}_\text{init} = \frac{1}{2}(\ket{x=d/2}+\ket{x=d/2+1})\otimes(\ket{1} -i\ket{-1}),
\end{equation}
\begin{figure}[t]
    \includegraphics[width=0.48\textwidth]{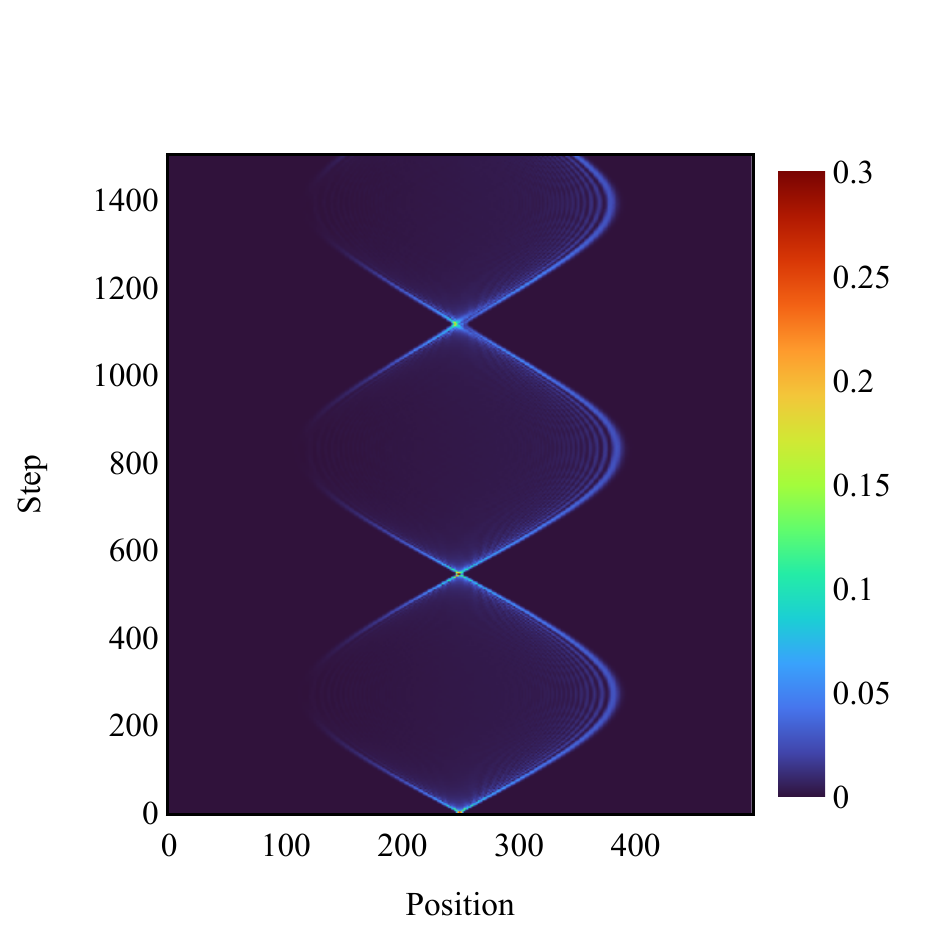}
    \caption{DTQW in weak stochastic electric gauge field evolution defined in Eq. \ref{ev2} of initial state $\ket{\Psi}=\frac{1}{\sqrt{2}}\ket{x=d/2}\otimes(\ket{1} -i\ket{-1})$ with $r = 0.9$, $\phi = \frac{2\pi}{d}$ and $d=501$. One can see the Bloch oscillations in the behavior of the particle even in the case of a random electric gauge field.}
    \label{dec2}
\end{figure}
with stochastic evolution defined above with $r=0.9$. The evolution averaged over 1000 realizations is shown in Fig. \ref{dec}. One can see that after a short time, the state stops behaving ballistic and instead becomes diffusive. As a result of decoherence that led to diffusive behavior of the walk, the probability distribution after $400$ steps is Gaussian (see Fig. \ref{state}). Thus even in a weak field regime, DTQW with a random (in time) gauge field defined by the generator given in Eq. \ref{phib} has a diffusive behavior. However, this shouldn't be surprising as it is common for random DTQW \cite{PhysRevLett.106.180403}.\ii
Let's now move on to the case of electric quantum walk. The electric gauge field will be generated by the following operator
\begin{equation}\label{phie}
    \Phi_E \ket{x,\pm} = e^{i\phi x}\ket{x,\pm}.
\end{equation}
Note that this phase shift has only spatial dependence. Again the evolution will be stochastic
\begin{align}
    \ket{\Psi(t+1)} = \begin{cases}
                   U\ket{\Psi(t)} & \text{with probability } (1-r),\\
                   U\Phi_{E}\ket{\Psi(t)} & \text{with probability } r.
              \end{cases}
    \label{ev2}
\end{align}
One could suspect that evolution again should be diffusive but surprisingly it is not the case. Even in DTQW with a weak stochastic electric gauge field, the particle undergoes the Bloch oscillations in the regime of $\phi\ll 1$ and $r\gg 0$ as one can see in the numerical simulations presented in Fig. \ref{dec2}.\ii
\begin{figure*}[t]
    \includegraphics[width=0.40\textwidth]{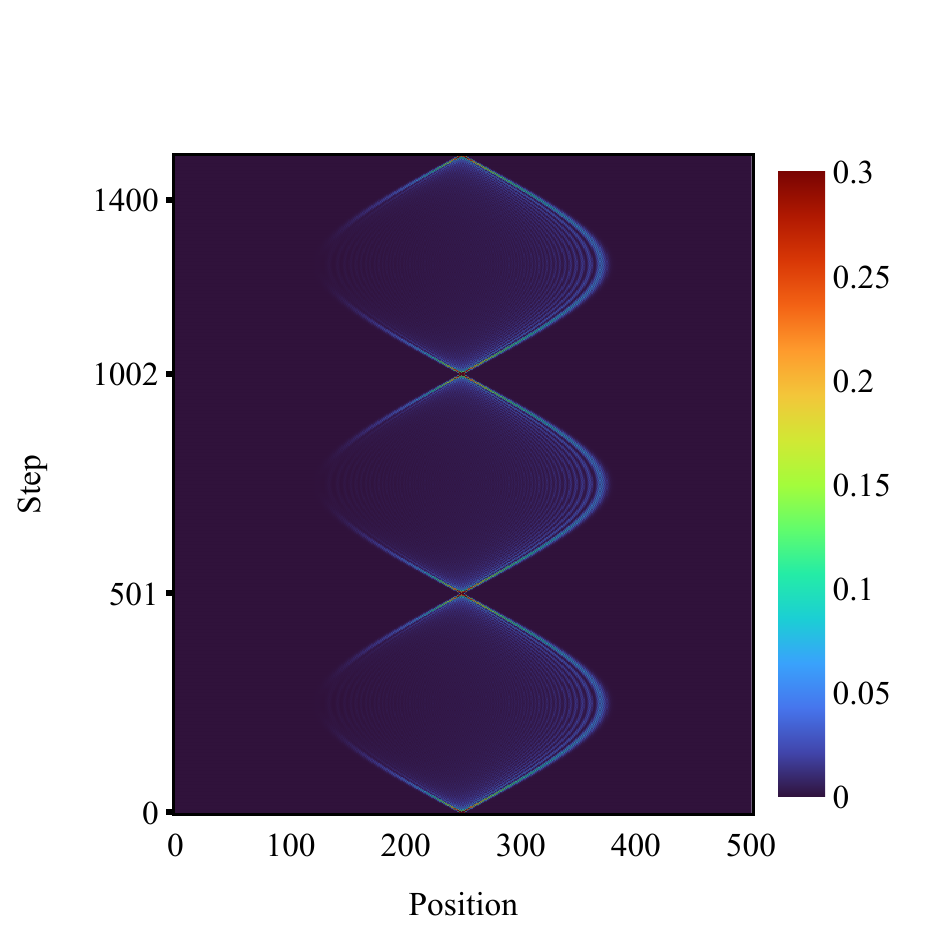}~~~~~~~~\includegraphics[width=0.40\textwidth]{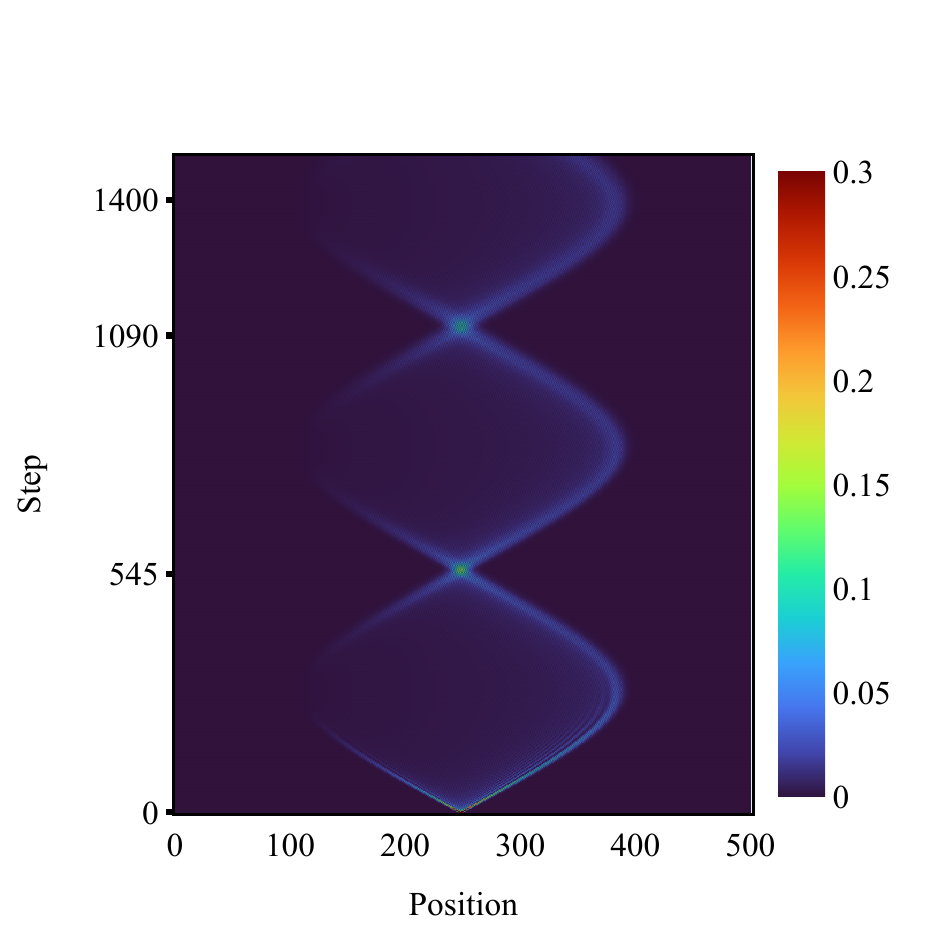}
    \caption{On the left side we plotted the evolution of standard electric quantum walk with $\phi = \frac{2\pi}{d}$ on a $d$ vertex cycle starting from initial state $\ket{\Psi}=\frac{1}{\sqrt{2}}\ket{x=d/2}\otimes(\ket{1} -i\ket{-1})$. As a result, we get standard Bloch oscillations with period $T = \frac{2\pi}{\phi} = 501$ steps. On the right side, we plotted the evolution of DTQW with stochastic electric gauge field (with $r = 0.9$) again with $\phi = \frac{2\pi}{d}$ on a $d$ vertex cycle averaged over 1000 realizations. Once again one can see the Bloch oscillations but this time with a longer period. It can be seen that the period of the oscillations is in agreement with our theorem $T \approx \frac{2\pi}{\phi(r+2(1-r)^2)} \approx 545$ steps.}
    \label{blochy}
\end{figure*}
To explain why the Bloch oscillations survive even in the random DTQW we will refer to the following commutation relations
\begin{gather}
    [\Phi_E,S] = i\phi\ \Omega_z  S\Phi_E,\\\label{ec}
    [\Phi_E,C] = 0,
\end{gather}
where 
\begin{equation}
    \Omega_z = I^{(d)} \otimes \sigma_z
\end{equation}
Using those relations one can derive
\begin{gather}\nonumber
    U\Phi_E = \Phi_E U + [U,\Phi_E] =  \Phi_E U +  U\Phi_E(i\phi\Omega_Z) =\\=\Phi_E U +  \Phi_E U(i\phi\Omega_z)+ U\Phi_E(-\phi^2),
\end{gather}
which due to the $\phi \ll 1$ we will approximate 
\begin{equation}\label{us}
    U\Phi_E \approx\Phi_E U e^{i\phi\Omega_Z}
\end{equation}
For sufficiently large $r$ we can assume that in the interval $l=(1-r)^{-1}$ steps there is one step at which the field is switched off. We call such a step defective. For example, let's consider the case of $r = 5/6$. An exemplary 6 step interval could look like this
\begin{equation}
    U_6 = U\Phi_EU\Phi_EU\Phi_EUU\Phi_EU\Phi_E.
\end{equation}
In this form, it is hard to tell what the evolution would look like. Our main goal is to get rid off defect on the third step of evolution. We can do that at the cost of the magnitude of the field by separating the field operators $\Phi_E = \Phi_E^{5/6}\Phi_E^{1/6}$. Next step is to leave the $\Phi_E^{5/6}$ and move the $\Phi_E^{1/6}$ to the defect. This swap is possible in the weak field regime ($\phi\ll 1$) where we can use Eq. \ref{us}.
\begin{gather}\label{u6}\nonumber
    U_6 =U\Phi_E^{5/6}e^{-i\frac{\phi}{6}\Omega_Z}U\Phi_E^{5/6}e^{-i2\frac{\phi}{6}\Omega_Z}U\Phi_E^{5/6}e^{-i3\frac{\phi}{6}\Omega_Z}\cdot\\ \cdot U\Phi_E^{5/6}U\Phi_E^{5/6}e^{i2\frac{\phi}{6}\Omega_Z}U\Phi_E^{5/6}e^{i\frac{\phi}{6}\Omega_Z}.
\end{gather}
We were able to erase the defected step at the cost of very small random in time rotations about the $z$ axis of the coin. Note that this rotation does not depend on position. This randomness is hidden in the prefactors of $\phi$. However this way easy to spot regularities arise which look promising in terms of averaging over the realizations. To this end let us rewrite the single step evolution operator in the following form
\begin{equation}
    U(t,j) = (I\otimes C) \ S \ \Phi_E^{r}\ e^{in_{t,j}\phi\Omega_Z},
\end{equation}
where
\begin{align}
    n_{t,j} = \begin{cases}
                   (i+1)(1-r) & \text{for } i < j,\\
                   0 & \text{for } i = j,\\
                   1-i(1-r) & \text{for } i > j.
              \end{cases}
\end{align}
the subscript $t$ denotes which step from the interval is it and $j$ denotes at which step was the defect. Averaging above over all possible $j$ we get
\begin{equation}
    \Bar{U}(t) = (I\otimes C) \ S \ \Phi_E^{r}e^{i\Bar{n}_{t}\phi\Omega_Z},
\end{equation}
where 
\begin{equation}
    \Bar{n}_t = -2(1-r)^2\left(t-\frac{(1-r)^{-1}-1}{2}\right).
\end{equation}
So the perturbation $e^{i\Bar{n}_{t}\phi\Omega_Z}$ linearly depends on time. However, as Cedzich and Werner had done in \cite{Cedzich2016} we can transform this time dependent evolution operator into only spatial dependent using the local gauge transformation.
\begin{equation}
    \Tilde{U} = V_t^\dagger \ \Bar{U}(t)\ V_{t-1},
\end{equation}
with 
\begin{equation}
    V_t = \sum_x \ket{x}\bra{x}\otimes e^{i\eta x t},
\end{equation}
where $\eta = -2\phi(1-r)^2$. So the time dependent $\Bar{U}(t)$ is gauge equivalent (up to phase that does not influence the evolution) to
\begin{equation}\label{tild}
    \Tilde{U} = (I \otimes C)\ S \ \Phi_E^{(r+2(1-r)^2)\phi}.
\end{equation}
The above unitary operator is just a standard electric DTQW evolution operator but with a changed phase. Since the period of Bloch oscillations is given by $T = \frac{2\pi}{\phi}$ the above described evolution should give rise to Bloch oscillations with a period
\begin{equation}
    T = \frac{2\pi}{\phi(r+2(1-r)^2)}.
\end{equation}
Thus for $r = 0.9$, we expect the period to be longer than in standard electric quantum walk, and in fact, the simulations we showed in Fig. \ref{blochy} reproduce that result.\ii
Getting back to Eq. \ref{tild} one can observe that when $r$ goes to $1$ the average evolution operator becomes
\begin{equation}
     \Tilde{U} = (I \otimes C)\ S \ \Phi_E^{r\phi}.
\end{equation}
Thus one can say that as the intuition suggests in the regime of $\phi \ll 1$ and with parameter $r$ close to $1$ the average evolution operator resembles the evolution operator of DTQW in an electric gauge field with magnitude $r\phi$.\ii
Note that the same line of reasoning could not be applied to the model we studied at the beginning. DTQW in weak stochastic gauge field dependent on both position and spin has nonvanishing commutator
\begin{equation}\label{bc}
    [\Phi_B,C] =  -2i\sin\theta \sum_x \ket{x}\bra{x}\otimes \sin(\phi x)\sigma_x.
\end{equation}
Because of the factor $\sin(\phi x)$ in the commutator the swapping procedure does not work. The crucial point leading to the striking difference in the weak field regime of both models is hidden in the commutation relations (Eq. \ref{ec} and Eq. \ref{bc}). In the second model, the evolution survives the randomness and does not turn classical because the field generating operator commutes with coin-toss and commutator $[\Phi_E,S]$ is proportional to $\phi$. Due to this property in the weak field regime where $\phi^2$ can be ignored, we were able to make use of Eq. \ref{us}. In the first model even in the weak field it is not an option. Note that due to the periodic boundary condition, the minimal value for $\phi$ is $\frac{2\pi}{d}$. So $sin(\phi x)$ spans values from $-1$ to $1$  regardless of $\phi$ being small. Thus we cannot move around the phase operators to get rid of the defect like we did in the second case Eq.\ref{u6}\ii
In summary, we compared two models of DTQWs in stochastic gauge fields. The first corresponds to a gauge field generated by a position and spin dependent phase factor, and the second corresponds to an electric quantum walk. Despite their common behavior in the strong field regime (diffusion), we found a striking difference in the weak field regime. Unexpectedly, the stochastic electric quantum walk survives the randomness and does not turn into diffusion. We managed to solve the model analytically by invoking the commutation relations of the model. We confirmed our observation in numerical simulations. Note also that electric quantum walks have been experimentally studied, for example, in setups like Cs atoms in optical lattices \cite{PhysRevLett.110.190601} or in coupled fiber loops \cite{regensburger2011zitterbewegung}. Thus, we believe that the presented phenomenon can be realized experimentally in similar setups.\ii
We want to express our gratitude to Antoni Wójcik for illuminating discussions. This research is supported by the Polish National Science Centre (NCN) under the Maestro Grant no. DEC-2019/34/A/ST2/00081.

\end{document}